\documentclass{llncs}

\begin{document}
\newcommand{\ehelp}{A} \newcommand{\ket}[1]{ | \, #1  \rangle}
\newcommand{\bra}[1]{ \langle #1 \,  |}
\newcommand{\proj}[1]{\ket{#1}\bra{#1}} \newcommand{\pket}[2]{ | \, #1
\rangle \otimes | \, #2  \rangle  } \newcommand{\pbra}[2]{ \langle #1
\,  | \otimes \langle #2 \,  |}
\newcommand{\pproj}[2]{\ket{#1}\bra{#1}\otimes\ket{#2}\bra{#2} }
\newcommand{\abs}[1]{ | \, #1 \,  |} \newcommand{\av}[1]{\langle
#1\rangle}

%
%\frontmatter          % for the preliminaries
%
\pagestyle{headings}  % switches on printing of running heads
%\addtocmark{Quantum Entanglement} % additional mark in the TOC

%\tableofcontents
%
\mainmatter              % start of the contributions

\title{Entanglement of Assistance} \author{David P. DiVincenzo\inst{1}
\and  Christopher A. Fuchs\inst{2} \and  Hideo Mabuchi\inst{2}\and
John A. Smolin\inst{1} \and Ashish Thapliyal \inst{3}  \and   Armin
Uhlmann\inst{4} }
\authorrunning{David P. DiVincenzo et al.}   
% abbreviated author list (for running head)
%
%%%% modified list of authors for the TOC (add the affiliations)
\tocauthor{David P. DiVincenzo (IBM Research), Christopher A. Fuchs
 (Cal.  Tech.), Hideo Mabuchi (Cal.Tech.), John A. Smolin (IBM
 Research), Ashish Thapliyal (UCSB),  Armin Uhlmann ( Universit\"at
 Leipzig)}
\institute{ IBM Research Division,  Yorktown Heights, NY 10598,
U.~S.~A. \email{divince/smolin@watson.ibm.com}\\ \and Bridge
Laboratory of Physics 12-33, California Institute of Technology,
Pasadena, CA 91125, U.~S.~A.  \email{cfuchs/hmabuchi@cco.caltech.edu}
\and Department of Physics, University of California at Santa Barbara,
Santa Barbara, CA 93106, U.~S.~A. \email{ash@physics.ucsb.edu} \and
Institut f\"ur Theoretische Physik, Universit\"at Leipzig,
Augustusplatz 10/11, D-04109 Leipzig,
Germany. \email{Armin.Uhlmann@itp.uni-leipzig.de} }

\maketitle              % typeset the title of the contribution

\begin{abstract}

The newfound importance of ``entanglement as a resource'' in quantum
 computation  and quantum communication behooves us to quantify it in
 as many distinct ways as possible. Here we explore a new
 quantification  of entanglement of a general (mixed) quantum state
 for a bipartite system, which we name {\it entanglement of
 assistance}. We show it  to be the maximum of the average
 entanglement over all ensembles consistent with the density matrix
 describing the bipartite state. With the help of lower and upper
 bounds we calculate entanglement of assistance for a few cases and
 use these results to  show the surprising property of
 superadditivity. We believe that this may throw some light on the
 question of additivity of {\it entanglement of  formation}. 
\end{abstract}
\section{Introduction}
Much of the preoccupation of classical information theory is in making
the correlation between two ends of a communication channel---that is,
between a sender and a receiver---as high as possible.  This is what
(classical) communication is about.  In contrast, this is only part of
the story for the fledgling field of {\it quantum\/} information
theory. The quantum world brings with it a new resource that senders
and receivers can share:  quantum entanglement, the stuff
Einstein-Podolsky-Rosen pairs and Bell-inequality violations are made
of.  This new resource, of all the things in quantum information
theory, is the most truly ``quantum'' of the lot.  It is a resource
because of the myriad ways in which it is starting to be used.  The
list of its applications already includes quantum key distribution
\cite{deu:eke:joz:mac:pop:96}, quantum-state teleportation
\cite{ben:bra:cre:joz:per:woo:93}, entanglement-enhanced classical
communication \cite{ben:fuc:smo:97}, error-correction for quantum
computers \cite{Got:97}, entanglement-assisted multi-party
communication \cite{cle:bur:97}, and better control of atomic
frequency standards  \cite{bol:ita:win:hei:96} among other things.
The list grows every day.

The newfound importance of ``entanglement as a resource'' behooves us
to quantify entanglement in as many distinct ways as possible.  To
this end, we explore a new quantification of the entanglement of a
general (mixed) quantum state for a bipartite system.  This quantity
we dub the {\it entanglement of assistance}.  It is something of a
dual notion to that of the {\it entanglement of formation\/} studied
by Bennett, DiVincenzo, Smolin and Wootters \cite{ben:div:smo:woo:96}.
It can be motivated in the following way.

Consider three players Alice, Bob, and Charlie, who jointly possess
many copies of a tripartite quantum system each described by the pure
state $|\Psi^{\rm\scriptscriptstyle ABC}\rangle$.  It follows that
Alice and Bob, considered in isolation, possess many copies of the
(generally mixed) state $\rho^{\rm\scriptscriptstyle AB}= {\rm
tr}_{\rm\scriptscriptstyle C} |\Psi^{\rm\scriptscriptstyle
ABC}\rangle\langle \Psi^{\rm\scriptscriptstyle ABC}|$.  Suppose now
that Alice and Bob need to use their shared quantum system for one of
the tasks above, teleportation for instance.  However, very
unfortunately, Charlie is not immediately available to pass his
systems over for their use.  Since their state may not be very pure or
entangled for that matter, they may be seriously impaired by this.
How might Charlie help in spite of his constraint?

One thing he can do is perform a measurement on each system that he
possesses so that Alice and Bob will have pure states conditioned on
his outcomes.  If he transmits the classical information so obtained,
they may be able to use it to their benefit.  For instance, if several
of his measurements reveal that Alice and Bob actually possess the
pure state $|\Psi^{\rm\scriptscriptstyle AB}\rangle$, then by the
result of Bennett, et al.~ \cite{ben:ber:pop:sch:96}, they can convert
these quantum states into maximally-entangled singlet states at a rate
of $ S(\rho^{\rm\scriptscriptstyle A})=- {\rm
tr}(\rho^{\rm\scriptscriptstyle A}\log_2 \rho^{\rm\scriptscriptstyle
A}) $, where $\rho^{\rm\scriptscriptstyle A}= {\rm
tr}_{\rm\scriptscriptstyle B} |\Psi^{\rm\scriptscriptstyle
AB}\rangle\langle \Psi^{\rm\scriptscriptstyle AB}|$.  In general then,
for a measurement that creates the states
$|\Psi_i^{\rm\scriptscriptstyle AB}\rangle$ with probabilities $p_i$,
i.e., an ensemble ${\cal E}=\{p_i,|\Psi_i^{\rm\scriptscriptstyle
AB}\rangle\langle \Psi_i^{\rm\scriptscriptstyle AB}|\}$, the average
rate for conversion to singlets will be
\begin{equation}
E({\cal E})=\sum_i p_i S(\rho_i^{\rm\scriptscriptstyle A})=-\sum_i
p_i\,{\rm tr}(\rho_i^{\rm\scriptscriptstyle A}\log_2
\rho_i^{\rm\scriptscriptstyle A})\;.
\end{equation}

Since Charlie is a friend of Alice and Bob, he should choose his
measurement so that this average rate is maximized.  By a theorem of
Hughston, Jozsa, and Wootters \cite{hug:joz:woo:93}  this maximization
can be taken over all possible ensembles $\cal E$ consistent with
$\rho^{\rm\scriptscriptstyle AB}$  in the sense that
$$
\rho^{\rm\scriptscriptstyle AB}= \sum_i
p_i|\Psi_i^{\rm\scriptscriptstyle AB}\rangle\langle
\Psi_i^{\rm\scriptscriptstyle AB}|\;.
$$

Therefore the resulting maximal rate is a number intrinsic to
$\rho^{\rm\scriptscriptstyle AB}$, depending upon no further details
of the state $|\Psi^{\rm\scriptscriptstyle ABC}\rangle$.  This is the
{\it entanglement of assistance} $A(\rho^{\rm\scriptscriptstyle AB})$
for $\rho^{\rm\scriptscriptstyle AB}$:
\begin{equation}
A(\rho^{\rm\scriptscriptstyle AB})= \max_{\cal E} E({\cal E})\;.
\label{eq:def:emax}
\end{equation}
This quantity seems to be dual to the entanglement of formation, which
is defined by
\begin{equation}
F(\rho^{\rm\scriptscriptstyle AB})= \min_{\cal E} E({\cal E})\;.
\end{equation}

In this paper, we demonstrate several properties possessed by
entanglement of assistance, calculating it explicitly for some
specialized cases.  We exhibit a general upper bound, a particular
upper bound specialized to the two-qubit case and a  lower bound for
diagonal  density matrices, on the entanglement of asistance. With the
aid of the last  two, we prove a fairly surprising property: the
entanglement of assistance is in some cases superadditive.  That is,
there exist density operators $\rho^{\rm\scriptscriptstyle AB}_{1,2}$
for which
\begin{equation}
A(\rho_1^{\rm\scriptscriptstyle AB}\otimes
\rho_2^{\rm\scriptscriptstyle AB})> A(\rho_1^{\rm\scriptscriptstyle
AB}) + A(\rho_2^{\rm\scriptscriptstyle AB})\;.
\end{equation}
This means that if Charlie performs entangled measurements on his
separate copies of the system, he can be more effective in helping
Alice and Bob than he would have been otherwise.

\section{Properties and bounds}
\par The average entanglement  of an ensemble is invariant under local
 unitary transformations and non-increasing under general local
 operations\footnote{Local operations are those that Alice and Bob can
 perform on their own part of the system with possible synchronization
 using classical messages. However, they may not send each other
 quantum systems nor can they come together and perform a joint
 quantum operation on the system.}  with classical communication
 \cite{ben:div:smo:woo:96,ben:ber:pop:sch:96}; so entanglement of
 assistance also has these properties.  In addition to this it is easy
 to see that since there is a maximization over all possible
 pure-state realizations of the density matrix, the entanglement of
 assistance $\ehelp$ is concave.  That is,
\begin{equation}
  \ehelp(\sum_i p_i\rho_i) \ge \sum_i p_i\ehelp(\rho_i) \enspace.
\end{equation}

 We note here that the entanglement of assistance is the least of
all concave functions coinciding on pure states with their partial
entropy i.e. entropy as seen from one side,  while entanglement of
fomation is the greatest convex function with that property. This may
be seen as a sign of the mentioned duality between the two.

   Since we do not have a general formula \footnote{We will present a
  formula for a non-trivial class of rank two density matrices in a
  forthcomming publication.}  for entanglement of assistance of a
  given density matrix, upper and lower bounds are of great
  importance. We have found a few bounds which have helped us
  calculate the entanglement of assistance in some cases and have been
  of immense value in proving its superadditivity.  

\subsection{Upper bounds}
  \par We have found an upper bound --- the entropic bound --- which
  works in general, and another upper bound --- the fidelity bound ---
  which works for the simplest case of one qubit on Alice's side and
  one on Bob's side. 

\subsubsection{Entropic bound:}
 The entanglement of assistance is never greater than the minimum of
the partial entropy as seen by Alice or Bob:
\begin{equation}
\ehelp(\rho) \le {\rm Min}[S({\rm tr_A}\rho),S({\rm tr_B
}\rho)]\enspace.
\label{eq:bound:entropic}
\end{equation}
To prove this  assume that $\ehelp(\rho)$ is achieved by the
pure-state  ensemble ${\cal E }=\{p_i,\Pi_i=\proj{\psi_i}\}$, then by
the  concavity of Von Neumann entropy $S$ we have,
\begin{equation}
 S({\rm tr_A} \rho) \ge \sum_i p_i S({\rm tr_A }\Pi_i) = \ehelp(\rho)
\enspace.
\end{equation}
The same inequality holds with trace over A replaced by trace over
B. Combining them we get the entropic bound.  

Since the Von Neumann entropy is additive and concave the bound is
also additive and  concave. Thus for states which saturate this bound
the entanglement of assistance must be additive. We can also see from
the results of section (\ref{sec:examples}) that the entropic bound is
zero if and only if  the entanglement of assistance  is zero. The
bound obviously agrees with entanglement of assistance for pure
states. It is easy  to show \footnote{We show this in a forthcomming
publication.}  that for the case of one qubit each on Alice's and
Bob's side, the entropic  bound is maximum (1 ebit) if and only if the
entanglement of assistance is also maximum.

\subsubsection{Fidelity bound:}
   For the case of one qubit each with Alice and Bob, we can get a
bound that is sometimes stronger than the entropic bound. It says that
the entanglement of assistance cannot be greater than the fidelity of
the density matrix relative to its complex conjugate in the magic basis
\footnote{The magic basis is: \\ $ \ket{e_1} =
\frac{\ket{00}+\ket{11}}{\sqrt{2}} $;  $ \ket{e_2} =
\frac{\ket{00}-\ket{11}}{-i\sqrt{2}} $; $ \ket{e_3} =
\frac{\ket{01}+\ket{10}}{-i\sqrt{2}} $;  $ \ket{e_4} =
\frac{\ket{01}-\ket{10}}{\sqrt{2}} $.}  i.e.,
\begin{equation}
   \ehelp(\rho) \le F(\rho,\tilde{\rho})
\label{eq:bound:fidelity}
\end{equation}
where the $ \tilde{\  } $ --- Hill-Wootters tilde --- represents
complex conjugation in the magic basis \cite{hil:woo:97}  and
$F(\rho,\sigma)= $ tr$\sqrt{\rho^{1/2}\sigma\rho^{1/2}}$ is  the
fidelity \cite{joz:94} that is, the square root of the transition
probability  \cite{uhl:76}. To prove the bound the main thing we need
is a classical inequality on the binary Shannon entropy
\cite{fuc:gra:97}.  It is,
\begin{eqnarray}
H_2(x) &=& -x\log_2 x-(1-x)\log_2(1-x)\\ &\le& 2\sqrt{x(1-x)}\;.
\end{eqnarray}
Equality is achieved in this if and only if $x=0,\frac{1}{2},1$.  The
other facts we need are  that for a pure state $\Pi_i$ in a
decomposition of $\rho$, the largest eigenvalue,
$\lambda_1(\rho_i^{\rm A}={\rm tr_B} \Pi_i)$, of its reduced density
operator is given in terms of the Hill-Wootters tilde by 
\begin{equation}
\lambda_1(\rho_i^{\rm A})=\frac{1}{2}\left[1+ \sqrt{1-{\rm
tr}(\Pi_i\tilde{\Pi}_i)}\,\right]
\end{equation}
and, that the fidelity function is doubly concave in its two arguments
and
\begin{equation}
F(\Pi_i,\tilde{\Pi}_i)=\sqrt{{\rm tr}(\Pi_i\tilde{\Pi}_i)}\;.
\end{equation}

Putting all this together:  for any ensemble ${\cal E}=\{p_i,\Pi_i\}$
consistent with $\rho$ average entanglement is 
\begin{eqnarray}
E({\cal E}) &=& \sum_i p_i S(\rho_i^{\rm A})\\ &=& \sum_i p_i
H_2\big(\lambda_1(\rho_i^{\rm A})\big)
\label{Fizzle}\\
&\le& 2\sum_i p_i \sqrt{\lambda_1(\rho_i^{\rm
A})\big(1-\lambda_1(\rho_i^{\rm A})\big)}
\label{Snubbton} \\
&=&  \sum_i p_i \sqrt{{\rm tr}(\Pi_i\tilde{\Pi}_i)}\\ &=& \sum_i p_i
F(\Pi_i,\tilde{\Pi}_i)
\label{SweatBack}\\
&\le& F(\rho,\tilde{\rho})\;.
\end{eqnarray}
The bound in Eq.~(\ref{eq:bound:fidelity}) follows immediately.

Note the conditions for saturating this bound.  In particular, in
going from Eq.~(\ref{Fizzle}) to (\ref{Snubbton}), we see that this
new bound can be achieved only if $\rho$ has a decomposition
consisting only of maximally entangled and completely unentangled
states.  Thus the bound can't be tight generally---consider simply a
$\rho$ that is pure but not itself maximally entangled.  However, we
will see in section (\ref{sec:superadditivity})  that this bound is
sometimes better than the entropic bound.

\subsection{Lower bounds}
\par  To get lower bounds we use the fact that the average
entanglement of any  pure-state realization of the density matrix
gives us a lower bound on the entanglement of assistance.  Using this
idea we can come up with lower bounds by making systematic pure-state
decompositions of the density matrix. For example, average
entanglement of the eigenvector decomposition of the density matrix is
a lower bound. 

 Applying the above idea to density matrices diagonal in a product
basis we get a lower bound which we name {\it diagonal lower bound}.
Consider  the case of only one qubit on either side so  that the
product basis  can be written as  $
\{|00\rangle,|01\rangle,|10\rangle,|11\rangle\} $.  The density matrix
in this product basis looks like,
\begin{equation} 
\rho = \left( 
\begin{array}{cccc}
\alpha_1 &  0 & 0 & 0 \\ 0 & \alpha_2 & 0 & 0 \\ 0 &  0 & \alpha_3 & 0
\\ 0& 0 & 0 & \alpha_4 \\
\end{array}
\right) 
\end{equation}
and the bound is,
\begin{equation}
\ehelp(\rho) \ge (\alpha_1+\alpha_4) H_2(\frac{\alpha_1}
{\alpha_1+\alpha_4}) + (\alpha_2+\alpha_3)
H_2(\frac{\alpha_2}{\alpha_2+\alpha_3}) \enspace . 
\label{eq:bound:diagonal}
\end{equation}
This follows from  the pure state realization of $\rho$:  $ {\cal E} =
\{p_i,|\psi_i\rangle \mid i=1,...,4 \} $, with
\begin{eqnarray}
\{p_i\} & = & \{(\alpha_1+\alpha_4)/2,(\alpha_1+\alpha_4)/2,
(\alpha_2+\alpha_3)/2,(\alpha_2+\alpha_3)/2 \} \nonumber \\
|\psi_1\rangle & = & (\sqrt{\alpha_1/(\alpha_1 +
\alpha_4)},0,0,\sqrt{\alpha_4/(\alpha_1+\alpha_4)}) \nonumber \\
|\psi_2\rangle & = & (\sqrt{\alpha_1/(\alpha_1 +
\alpha_4)},0,0,-\sqrt{\alpha_4/(\alpha_1+\alpha_4)}) \nonumber \\
|\psi_3\rangle & = & (0,\sqrt{\alpha_2/(\alpha_2 +
\alpha_3)},\sqrt{\alpha_3/(\alpha_2+\alpha_3)},0) \nonumber \\
|\psi_4\rangle & = & (0,\sqrt{\alpha_2/(\alpha_2 +
\alpha_3)},-\sqrt{\alpha_3/(\alpha_2+\alpha_3)},0) \enspace .
\end{eqnarray}
Numerical results using optimization algorithms suggest that this is
indeed the exact formula for entanglement of assistance for these
density matrices. Also when this result saturates any of the upper
bounds we know that it is the answer.

  The bound can easily be generalized to higher dimensions
\footnote{This will be presented in a forthcomming publication.
}. Another thing to note here is that off-diagonal terms between
$\{\ket{00},\ket{11}\}$ and $\{\ket{01},\ket{10}\}$ don't change the
lower bound because we can still use a similar pure state
decomposition as before, with slightly modified probabilities.

\section{Examples}
\label{sec:examples}

\par Let us start by characterizing the states that have zero
entanglement of assistance. We find that there is a very simple
characterization of these states as, states whose density matrix  is
pure on at least one side. That is,  either  $ \rho =
|\psi_{A}\rangle\langle\psi_{A}| \otimes \rho_{B} $ or $\rho =
\rho_{A} \otimes |\psi_{B}\rangle\langle\psi_{B}| $.   

To prove this result first note that since the entropic  (upper) bound
(\ref{eq:bound:entropic}) is zero for such states, these states have
zero entanglement of assistance. Let us now prove that only such
states can have zero  entanglement of assistance.  Since entanglement
of assistance is zero, every valid pure-state decomposition of $\rho$
must consist exclusively of product states.  (For otherwise the
average entanglement of that ensemble will be necessarily greater then
zero.)  So let us focus on an eigendecomposition of $\rho$: this is an
ensemble ${\cal E}=\{\lambda_i,
|\alpha_i\rangle|\beta_i\rangle\}$. (The $\lambda_i\ne0$ are the
probabilities for the associated states.)

Now, by the Hughston-Jozsa-Wootters theorem \cite{hug:joz:woo:93}, any
``unitary reshuffling'' of this ensemble is also a valid ensemble.
Therefore let us consider a new ensemble ${\cal E}^\prime=\{p_i,
|\psi_i\rangle\}$ that is exactly the same as the old one, except in
the first two elements.  Namely,
\begin{eqnarray}
\sqrt{p_1}|\psi_1\rangle &=&
\cos\theta\sqrt{\lambda_1}|\alpha_1\rangle|\beta_1\rangle+
\sin\theta\sqrt{\lambda_2}|\alpha_2\rangle|\beta_2\rangle \\
\sqrt{p_2}|\psi_2\rangle &=&
\sin\theta\sqrt{\lambda_1}|\alpha_1\rangle|\beta_1\rangle-
\cos\theta\sqrt{\lambda_2}|\alpha_2\rangle|\beta_2\rangle\;.
\end{eqnarray}

Because the ensemble $\cal E$ is an eigenensemble, we must have either
$\langle\alpha_1|\alpha_2\rangle=0$ or
$\langle\beta_1|\beta_2\rangle=0$ or both.  First note that it cannot
be the case that both inner products vanish.  For otherwise, there
will be at least one value of $\theta$ for which $|\psi_1\rangle$ and
$|\psi_2\rangle$ will be entangled states.  Consequently, let us look
at one of the other possibilities:  namely, $|\alpha_1\rangle$ and
$|\alpha_2\rangle$ are othogonal, but
$\langle\beta_1|\beta_2\rangle\ne0$.  Then we must have
$|\beta_1\rangle=e^{i\phi}|\beta_2\rangle$.  For otherwise,
again---for at least one value of $\theta$---both $|\psi_1\rangle$ and
$|\psi_2\rangle$ will be entangled states. The argument goes
analogously if $|\beta_1\rangle$ and $|\beta_2\rangle$ are othogonal.
That is to say, $|\psi_1\rangle$ and $|\psi_2\rangle$ must indeed be
product states but with either identical left factors or identical
right factors.

Therefore, just to say it again (but now ignoring phases, since they
could have been in the ``unitary reshuffling'' anyway), we must have
either $|\alpha_1\rangle=|\alpha_2\rangle$ or
$|\beta_1\rangle=|\beta_2\rangle$.  Locking one of these possiblities
in, we can go through exactly the same argument for similar ``unitary
reshufflings'' of other pairs of eigenstates.  Consistency then gives
that we must find either
\begin{equation}
\rho = |\psi_{A}\rangle\langle\psi_{A}|  \otimes \rho_{B}
\quad\mbox{or}\quad \rho =  \rho_{A} \otimes
|\psi_{B}\rangle\langle\psi_{B}|\;.
\end{equation}
This completes the proof.

Let us now look at the states with the maximum value \footnote{ The
maximum value is ${\rm Log_2}({\rm Min[N_A,N_B]})$ ~ebits. Here, ${\rm
N_A}$ and ${\rm N_B}$ are the dimensions of the Hilbert spaces on
Alice's and Bob's side respectively.} for entanglement of
assistance. This is the set of all possible mixtures of maximally
entangled pure-states. For the simplest non-trivial case of  one qubit
on either side, this is equivalent to the set of `T' states of
Horodecki et al. \cite{hor:hor:96,hor:hor:hor:96} or equivalently the
set of real \footnote{upto a phase.} density matrices in the magic
basis \cite{hil:woo:97}. 
   
Next we consider some non-extremal examples.  Consider states like
$\rho=$ Diagonal[$\alpha,0,0,1-\alpha$] in a product basis. The
entropic bound (\ref{eq:bound:entropic}) says  $\ehelp(\rho) \le
H_2(\alpha)$ and the diagonal lower bound (\ref{eq:bound:diagonal})
gives us $\ehelp(\rho)\ge H_2(\alpha)$ so that
$\ehelp(\rho)=H_2(\alpha)$.  

Let us look at a more complicated example, say $\rho=$
Diagonal[1/3,1/3,1/3,0] in a product basis.  The diagonal bound
(\ref{eq:bound:diagonal}) gives us $\ehelp(\rho) \ge 2/3 $. For the
fidelity bound, we first calculate $\tilde{\rho} =
$Diagonal[0,1/3,1/3,1/3] and  by the fidelity bound
(\ref{eq:bound:fidelity}) we see that $\ehelp(\rho) \le 2/3 $. Thus
$\ehelp(\rho) =2/3$. Notice that the entropic bound for this case is
just $H_2(1/3)\approx 0.918$~ebits. Since it does not saturate the
entropic bound this state is a good candidate for superadditivity.
Let us turn to that next.

\section{Superadditivity}
\label{sec:superadditivity}
\par It has been speculated (and is indicated numerically for a few
cases) that the entanglement of formation is additive
\cite{smo:woo:97}. This property is very important if it is to be used
as  a measure of entanglement.  Since entanglement of assistance looks
dual to entanglement of formation it comes as a bit of a surprise then
that the entanglement of formation is superadditive. By this we mean
\begin{equation}
   \ehelp(\rho_1^{\rm AB} \otimes \rho_2^{\rm AB}) >
   \ehelp(\rho_1^{\rm AB}) + \ehelp(\rho_2^{\rm AB})
\label{eq:superadddef}
\end{equation}
for some  density matrices  $\rho_1^{\rm AB}$ and $\rho_2^{\rm AB}$.
 Since the entropic bound is additive, superadditivity is possible
 only for the states which do not saturate the bound. 

The density matrix $\rho=$Diagonal[1/3,1/3,1/3,0] which we have just
 seen in section (\ref{sec:examples}) has an entanglement of
 assistance $\ehelp = 2/3$. Recall that this state does not saturate
 the (additive) entropic bound. So it can show superadditivity and in
 fact it does!  If $\ehelp$ were additive: $\ehelp(\rho \otimes \rho)=
 2 \ehelp(\rho) =4/3 $.  However $\rho \otimes \rho$ can be realized
 as an ensemble ${\cal E}$ which has an average entanglement $E
 \approx 1.5506$. (This ensemble is shown in detail in appendix
 \ref{ap:superadd}). This being a lower bound on $\ehelp(\rho \otimes
 \rho) $ it shows that entanglement of assistance  is superadditive.  

Superadditivity of entanglement of assistance means that if Alice,
Bob, and Charlie have two copies of the same system then Charlie can
give Alice and Bob more entanglement if he makes an entangled
measurement on the  two copies of the system. 

Since entanglement of assistance and entanglement of formation are
dual to each other in the sense of replacing a maximization by a
minimization, we would expect the superadditivity of the entanglement
of assistance to tell us something about the additivity of the
entanglement of formation.  We have not been able to see a connection
as yet but we note here a result \cite{ben:nie:tha:woo:97},
\begin{equation}
A(\rho) - F(\rho) \le S(\rho) - \abs{S({\rm tr_A} \rho)- S({\rm
tr_B}\rho)} \enspace,
\end{equation}
which connects entanglement of assistance and entanglement of
formation.  Thus in addition to its own physical  significance,
entanglement of assistance may turn out to be a useful tool to study
the problem of additivity of the entanglement of formation.

\section{Conclusions} 
In this paper we have introduced a new measure of entanglement ---
entanglement of assistance --- and studied its properties and found
upper and lower bounds for it. We have proved that it is superadditive
meaning that, entangled measurements by Charlie give more entanglement
to Alice and Bob. We also note here that the superadditivity of
entanglement of assistance may throw some light upon the additivity
question  of the entanglement of formation.
	
\section{Acknowledgments} 
Part of this work was completed during the 1997 Elsag-Bailey -- I.S.I.
Foundation research meeting on quantum computation.  We would like to
thank Charles H. Bennet, William K. Wootters and Barbara M. Terhal for
illuminating discussions. DPD and JAS would like to thank the Army
Research Office for support. CAF was supported by a Lee A. DuBridge
Fellowship and by DARPA through the Quantum Information and Computing
(QUIC) Institute administered by the US Army Research Office. AT would
like to thank The NSF Science and Technology Center for Quantized
Electronics Structures, Grant \#DMR 91-20007, for support to attend
this conference and, IBM Research Division for supporting his summer 
visit during which this work was initiated. He would
also like to thank Prof. David Awschalom for his invaluable support,
without which, it would have been impossible to work in this exciting
field.

\appendix
\section{Superadditivity}
\label{ap:superadd}
Here we look at the ensemble which gives us an average entanglement
more than the additive value of 4/3 for the density matrix discussed
is section (\ref{sec:superadditivity}).  The ensemble written in the
product basis  $\{ \ket{00_{\rm A}00_{\rm B}},\ket{00_{\rm A}01_{\rm
B}},..., \ket{11_{\rm A}11_{\rm B}} \}$ is, $ {\cal E} =
\{p_i,|\psi_i\rangle, i=1,..,12 \}$ where,
\begin{eqnarray}
  \{p_i\} & = &
           (\alpha_1,\alpha_1,\alpha_1,\alpha_1,\alpha_1,\alpha_1,
           \alpha_2,\alpha_2,\alpha_2,\alpha_2,\alpha_2,\alpha_2)\nonumber
           \\ |\psi_1\rangle & = & ( 0,0,0,a,0,0,b
           e^{i\pi/3},0,0,be^{-i\pi/3},0,0,a,0,0,0 )  \nonumber \\
           |\psi_2\rangle & = & ( 0,0,0,a,0,0,-b e^{i\pi/3},0,0,-b
           e^{-i\pi/3},0,0,a,0,0,0 ) \nonumber \\ |\psi_3\rangle & = &
           ( 0,0,0,a,0,0,b e^{i2\pi/3},0,0,b e^{-i2\pi/3},0,0,-a,0,0,0
           )  \nonumber \\ |\psi_4\rangle & = & (0,0,0,a,0,0,-b
           e^{i2\pi/3},0,0,-b e^{-i2\pi/3},0,0,-a,0,0,0)  \nonumber \\
           |\psi_{5}\rangle & = & (d,0,0,0,0,0,b,0,0,b,0,0,0,0,0,0)
           \nonumber \\ |\psi_{6}\rangle & = &
           (d,0,0,0,0,0,-b,0,0,-b,0,0,0,0,0,0) \nonumber \\
           |\psi_7\rangle & = & ( 0,c,0,0,0,0,b,0,c,0,0,0,0,0,0,0 )
           \nonumber \\ |\psi_8\rangle & = & ( 0,c,0,0,0,0,b
           e^{i2\pi/3},0,c e^{-i2\pi/3},0,0,0,0,0,0,0 )  \nonumber \\
           |\psi_9\rangle & = & ( 0,c,0,0,0,0,b e^{i4\pi/3},0,c
           e^{-i4\pi/3},0,0,0,0,0,0,0 )  \nonumber \\
           |\psi_{10}\rangle & = & ( 0,0,c,0,c,0,0,0,0,b,0,0,0,0,0,0 )
           \nonumber \\ |\psi_{11}\rangle & = & ( 0,0,c,0,c
           e^{i2\pi/3},0,0,0,0,b e^{-i2\pi/3},0,0,0,0,0,0 ) \nonumber
           \\ |\psi_{12}\rangle & = & ( 0,0,c,0,c
           e^{i4\pi/3},0,0,0,0,b e^{-i4\pi/3},0,0,0,0,0,0) \nonumber 
\end{eqnarray} 
and,
\begin{eqnarray}
  a & = & \left( 6-\frac{24}{5+\sqrt{7}} \right) ^{-\frac{1}{2}}
  \approx  0.5912 \nonumber \\ 
  b & = & \left( 9-\frac{18}{5+\sqrt{7}} \right)^{-\frac{1}{2}} 
  \approx 0.3879 \nonumber \\ 
  c & = & \left( \frac{5+\sqrt{7}}{18} \right)^{\frac{1}{2}} 
  \approx 0.6517 \nonumber  \\ 
  d & = & \left(3-\frac{12}{5+\sqrt{7}} \right) ^{-\frac{1}{2}} 
  \approx  0.8361 \nonumber \\
  \alpha_1 & = & \frac{1}{6}-\frac{2}{3(5+\sqrt{7})} 
  \approx 0.0795 \nonumber \\
  \alpha_2 & = & \frac{2}{3(5+\sqrt{7})} 
  \approx  0.0872.
\end{eqnarray}

The entanglements for the pure-states forming the ensemble $ {\cal E}
$  are,
\begin{equation}
 \{E_i\} =
(E_\alpha,E_\alpha,E_\alpha,E_\alpha,E_\beta,E_\beta,E_\gamma,
E_\gamma,E_\gamma,E_\gamma,E_\gamma,E_\gamma)
\end{equation}
where,
\begin{eqnarray}
 E_\alpha & = & H_2(a^2,a^2,b^2,b^2) 
\approx  1.8824 \;\mathrm{ebits} \enspace , \nonumber \\ 
 E_\beta & = & H_2(d^2,b^2,b^2,0) 
 \approx 1.1834 \;\mathrm{ ebits} \enspace , \nonumber \\
 E_\gamma & = & H_2(c^2,b^2,c^2,0) 
\approx 1.4605 \;\mathrm{ ebits} \enspace , \nonumber 
\end{eqnarray}
and  average entanglement of this ensemble is,
\begin{equation}
E({\cal E})=4\alpha_1 E_\alpha + 2 \alpha_1 E_\beta + 6 \alpha_2
E_\gamma  \approx 1.5506 \;\mathrm{ebits}.
\end{equation}

To find this ensemble we use the following procedure: First, we use a
numerical optimization program to find the ensemble which  has the
maximum average entanglement. Then using the structure of this
ensemble, guessing the right phases and using the constraint that the
ensemble must give us  $\rho  \otimes \rho$ we get a quadratic
equation which we solve to find the  values for $a$,  $b$, $c$, $d$,
$\alpha_1$ and $\alpha_2$. 

Note that according to our numerical results, the  ensemble presented
here achieves the maximum average entanglement consistent  with the
density matrix. This  suggests that  this is the entanglement of 
assistance for it.


\begin{thebibliography}{}


% using this for quantum key distribution ref.
%``Quantum Privacy Amplification and the Security of
%Quantum Cryptography Over Noisy Channels,''
\bibitem{deu:eke:joz:mac:pop:96} C.H. Bennett and G. Brassard ``Quantum
Cryptography: Public Key Distribution and Coin Tossing'', Proceedings
of IEEE International Conference on Computers Systems and Signal
Processing, Bangalore India, December 1984, pp 175-179.; D.~Deutsch,
A.~Ekert, R.~Jozsa, C.~Macchiavello, S. Popescu,  and A.~Sanpera,
Phys.\ Rev.\ Lett.\ {\bf 77}, 2818 (1996).

%Quantum Teleportation
\bibitem{ben:bra:cre:joz:per:woo:93} C.~H. Bennett, G.~Brassard,
C.~Cr\'{e}peau, R.~Jozsa, A.~Peres, and W.~K. Wootters, 
%``Teleporting an Unknown Quantum State via Dual
%Classical and {E}instein-{P}odolsky-{R}osen Channels,'' 
Phys.\ Rev.\ Lett.\ {\bf 70}, 1895 (1993).

%Entangled enhanced classical communication
\bibitem{ben:fuc:smo:97} C.~H. Bennett, C.~A. Fuchs, and J.~A. Smolin,
``En\-tangle\-ment-Enhanced Classical Communication on a Noisy Quantum
Channel,'' in {\sl Quantum Communication, Computing and Measurement},
edited by O.~Hirota, A.~S. Holevo, and C.~M. Caves (Plenum, New York,
1997).

%Error-correction for quantum computers using entanglement
\bibitem{Got:97}
P.~W.~Shor, Phys. Rev. {\bf A 52}, 2493 (1995);
 D.~Gottesman,
% {\sl Stabilizer Codes and Quantum Error Correction},
Ph.~D. Thesis, California Institute of Technology, 1997, LANL e-print
{\tt quant-ph/9705052}.

%Entanglement-assisted multi-party communication 
\bibitem{cle:bur:97} R.~Cleve and H.~Buhrman, 
%``Substituting Quantum Entanglement for Communication,''  
 LANL e-print {\tt quant-ph/9704026}.

%Better control of atomic freq. standards using entangled states.
\bibitem{bol:ita:win:hei:96} J.~J. Bollinger, W.~M. Itano,
D.~J. Wineland, and D.~J. Heinzen,
%``Optimal Frequency Measurements with Maximally Correlated States,''
Phys.\ Rev.\ A {\bf 54}, R4649 (1996).

% Mixed State Entanglement and Quantum Error Correction
\bibitem {ben:div:smo:woo:96}  C.\  H.\  Bennet, D.\  P.\  DiVincenzo,
J.\ A.\  Smolin, W.\  K.\  Wootters, Phys.\  Rev.\ A {\bf54},
3824(1996), LANL e-print {\tt quant-ph/9604024}.

% Concentrating Partial Entanglement by Local Operations. 
% I've read this one.
\bibitem{ben:ber:pop:sch:96} C.\ H.\ Bennett, H.\ J.\ Bernstein, S.\
Popescu, B.\ Schumacher,  Phys.\ Rev.\ {\bf A 53}, 2046(1996),
LANL e-print {\tt quant-ph/9511030}.

%This one shows the unitary shuffling of an ensemble realizing 
%a density matrix is another ensemble that realizes it. 
% All ensembles can be characterized in terms of an eigenensemble
%and unitary shuffling.
\bibitem {hug:joz:woo:93}  L.~P.~Hughston, R.~Jozsa, W.\ K.\ Wootters,
Phys.\ Lett.\ A {\bf 183}, 14 (1993).  

%Entanglement of a Pair of Quantum Bits. 
%Tilde defined in Magic basis.
%Generalized bell states is equivalent to T states and to real
%(upto a phase) density matrices in the bell basis. This is the set of 
%all mixtures of the Bell states.
\bibitem{hil:woo:97}  S.~Hill, W.~K.~Wootters, Phys.~Rev.~Lett. {\bf
78} ,5022 (1997),  LANL e-print {\tt quant-ph/9703041}.

% fidelity for mixed states
\bibitem{joz:94}
R. Jozsa, J. Mod. Opt. {\bf 41}, 2315 (1994).

%Transition probability for mixed states (= Square of Fidelity)
\bibitem{uhl:76}
A.~Uhlmann, Rep.~Math.~Phys. {\bf 9}, 273 (1976).
%((The ``Transition Probability'' in the State Space of a *-Algebra))

%The inequality $H_2(x) \le 2\sqrt{x(1-x)}$ proved by graphing.
%Cryptographic Distinguishability Measures for Quantum Mechanical States
\bibitem{fuc:gra:97}
C. A. Fuchs and J. van de Graaf,  LANL e-print {\tt quant-ph/9712042}.

% T-states defined maybe and equivalence to general bell mixtures.
\bibitem{hor:hor:96}
R.~Horodecki, M.~Horodecki, Phys.~Rev.~A {\bf 54}, 1838 (1996).

%T-states defined maybe and equivalence to general bell mixtures.
\bibitem{hor:hor:hor:96}  
R.~Horodecki, M.~Horodecki, P.~Horodecki, Phys. Lett. A {\bf 222}, 
21 (1996).

%Evidence for additivity of E_formation.
%
\bibitem{smo:woo:97}
J.~A.~Smolin, private communication;
W.~K. Wootters, 
%``Entanglement of Formation of an Arbitrary State
%of Two Qubits,''
 to appear in Phys.\ Rev.\ Lett., LANL e-print {\tt quant-ph/9709029}.

\bibitem{ben:nie:tha:woo:97}
M.~Nielsen, private communication; The same result was independently 
conjectured  in a discussion with C.~H.~Bennett and W.~K.~Wootters. 

%Quantum Superdense Coding.
%\bibitem{ben:wie:92}
%C.~H.~Bennett, S.~J.~Wiesner, Phys. Rev. Lett. {\bf 69}, 2881 (1992).

\end{thebibliography}
\end{document}